\documentstyle[12pt,epsf]{article}

\begin{document}

\title{Hypersensitivity to Perturbations in the
Quantum Baker's Map\thanks{Submitted to Physical Review Letters}}

\author{R\"udiger Schack$^{^{\hbox{\tiny(a)}}}$ and
        Carlton M. Caves$^{^{\hbox{\tiny(a,b)}}}$ \\
  \\
$^{\hbox{\tiny(a)}}${\it Department of Physics and Astronomy,
University of New Mexico} \\
{\it Albuquerque, NM~87131} \\
$^{\hbox{\tiny(b)}}${\it Santa Fe Institute,
1660 Old Pecos Trail, Suite A,
Santa Fe, NM~87501}}

\date{January 14, 1993}

\maketitle

\begin{abstract}
We analyze a randomly perturbed quantum version of the baker's
transformation, a prototype of an area-conserving chaotic map. By
numerically simulating the perturbed evolution, we estimate the
information needed to follow a perturbed Hilbert-space vector in
time. We find that the Landauer erasure cost associated with this
information grows very rapidly and becomes much larger than the
maximum statistical entropy given by the logarithm of the dimension of
Hilbert space. The quantum baker's map thus displays a
hypersensitivity to perturbations that is analogous to behavior found
earlier in the classical case. This hypersensitivity characterizes
``quantum chaos'' in a way that is directly relevant to statistical
physics.
\end{abstract}

\section*{}

Great progress has been made in studying manifestations of chaos
in quantum systems \cite{QuantumChaos}, yet there still is
controversy as to whether quantum chaos exists at all
\cite{Ford1991,Peres1992}. A chief reason for this is that the most
important characteristic of classical chaotic systems---exponential
divergence of trajectories starting at arbitrarily close initial
points in phase space---is absent from quantum systems simply because
the existence of a quantum scale makes meaningless the concept of two
arbitrarily close points in phase space.

Any attempt to find exponential divergence of trajectories of
Hilbert-space vectors founders immediately, because the linear
Schr\"odinger equation, with its unitary evolution, preserves
Hilbert-space inner products. Yet the unitary linear evolution of the
Schr\"odinger equation must be irrelevant to the issue of quantum
chaos \cite{Ford1991}, since any Hamiltonian classical chaotic system
can be described by an analogous area-conserving linear Liouville
equation. Two probability distributions in classical phase space---we
call these {\em phase-space patterns\/}---that are initially close
together (in terms of an overlap integral) stay close together
forever, because the Liouville equation is area conserving. This leads
to the question we address in the present article: Are there
manifestations of Hamiltonian classical chaos in the Liouville
equation that are also present in quantum mechanics?

In an earlier paper \cite{Schack1992} we analyzed a prototype of an
area-conserving chaotic map, the baker's transformation
\cite{Arnold1968}, in the Liouville representation; i.~e., we focused
on phase-space patterns instead of on single trajectories. Our
analysis was guided by the question \cite{Caves1993} of how available
work decreases with time when the baker's map is subjected to
area-conserving random perturbations. An area-conserving abstract
mapping corresponds to an energy-conserving phase-space system,
so we identify two negative contributions to free energy. The
conventional one is ordinary entropy, which measures how incomplete
knowledge about a system reduces our ability to extract work. The
other contribution arises from Landauer's principle
\cite{Landauer1961,Landauer1988} that there is an unavoidable energy
cost of $k_BT\ln2$ connected with the erasure of one bit of
information. It follows from Landauer's principle that the
information, quantified by algorithmic information \cite{Chaitin1987},
needed to give a complete description of a system state also reduces
the amount of available work and thus should make a further negative
contribution to free energy \cite{Zurek1989a,Zurek1989b}.

In our earlier paper \cite{Schack1992} we compared two strategies for
preserving the ability to extract work from a system. The first
strategy is to keep track of the perturbed phase-space pattern in {\it
fine-grained\/} detail, in an attempt to preserve the work inherent in
the initial condition. The second strategy, which we call {\it coarse
graining}, is to average over the perturbation and to put up with the
resulting ordinary entropy increase. We found, for the perturbed
baker's map, that the information needed to implement the first
strategy is overwhelmingly larger than the entropy increase of the
second strategy. This means that the free-energy cost of tracking the
perturbed pattern in fine-grained detail is enormous and far greater
than the cost of the entropy increase that results from coarse
graining. We conjecture that this {\em hypersensitivity to
perturbations\/} is a general feature of perturbed classical chaotic
systems, and we regard it as the desired manifestation of classical
chaos in the Liouville equation.

In the present paper we compare the two strategies for preserving work
in the case of a quantum system, a quantum version of the baker's
map \cite{Balazs1989}. Using numerical simulation, we find essentially
the same behavior as in the classical case, as was suggested using
heuristic arguments in Ref.~\cite{Caves1993}. The quantum baker's map
displays hypersensitivity to perturbations and thus can be said to
exhibit quantum chaos.

The concept of algorithmic information has been used before to
investigate quantum chaos \cite{Ford1991,Weigert1991}. If one defines
a chaotic system as one where the algorithmic information needed to
predict a single (unperturbed) trajectory grows linearly with time (or
number of steps) \cite{Ford1983}, then there is classical chaos,
but no quantum chaos \cite{Ford1991}. Our approach, by focusing on
patterns in phase space instead of trajectories, uses a framework
where classical and quantum mechanics can be treated on
analogous footings. Moreover, since Landauer's principle gives
information an explicit physical meaning by connecting it to available
work, our characterization in terms of hypersensitivity to
perturbations is directly relevant to statistical physics.

The classical baker's transformation maps the unit square $0 \leq
q,p \leq 1$ onto itself according to
\begin{equation}
f:(q,p) \longmapsto ( 2q-[2q], (p+[2q])/2 ) \;,
\label{cmap}
\end{equation}
where the square brackets denote the integer part.  There is no unique
way to quantize a classical map. Here we adopt a quantized baker's map
introduced by Balazs and Voros \cite{Balazs1989} and put in more
symmetrical form by Saraceno \cite{Saraceno1990}. Position
and momentum space are discretized, placing the lattice sites at
half-integer values
$q_j=p_j=(j+1/2)/2N$ for $j=0,\ldots,2N-1$.
The dimension $2N$ of Hilbert space is assumed to be even. For
consistency of units, let the quantum scale on phase space be
$2\pi\hbar=1/2N$. Position and momentum basis kets are denoted by
$|q_j\rangle$ and $|p_j\rangle$. A transformation
between these two bases is performed by the operator $G_{2N}$, defined
by the matrix elements
\begin{equation}
(G_{2N})_{jk} = \langle p_j|q_k\rangle = \sqrt{2\pi\hbar} \;
e^{-ip_jq_k/\hbar} \;.
\label{Fourier}
\end{equation}
The quantum baker's map is now defined by the matrix
\begin{equation}
B = G_{2N}^{-1} \left( \begin{array}{cc}
                       G_N & 0 \\ 0 & G_N
                       \end{array} \right) \;,
\label{qmap}
\end{equation}
where, as throughout this article, matrix elements and vector
coordinates are given relative to the position basis.

The perturbation operator we use is constructed to resemble the type
of perturbation used in our previous work \cite{Schack1992} on the
classical baker's map. We partition phase space into an even number
$2N_c=2N/w_c$ of congruent {\em perturbation cells}, where $2N_c$ and
$w_c\geq2$ are integral divisors of $2N$.  In the following we use
perturbation cells that are vertical stripes extending over the entire
$p$ range. Then each perturbation cell contains $w_c$ $q$-eigenstates.
A perturbation operator that perturbs each perturbation cell
independently has the form of a matrix with zero elements everywhere
except for $2N_c$ square blocks of size $w_c$ on the diagonal. We
choose these square blocks so that they correspond to a shift in the
$p$ direction in a $w_c$-dimensional subspace. Let the momentum shift
in the $n$th perturbation cell ($n=0,\ldots,2N_c-1$) be $\alpha i_n$,
where the real number $\alpha$ is the magnitude of the momentum shift
and $i_n\in\{-1,1\}$. The symmetry condition $i_{2N_c-n-1}=i_n$
($n=0,\ldots,N_c-1$) avoids rapid oscillations and thus ensures
similarity to the classical case. The perturbation operator
$U_{\alpha;i_0,\ldots,i_{N_c-1}}$ is defined by
\begin{equation}
(U_{\alpha;i_0,\ldots,i_{N_c-1}})_{jk} = \delta_{jk}e^{i2\pi\phi_k}\;,
\label{pert}
\end{equation}
where $\phi_0=0$, $\phi_k=\phi_{k-1} + \alpha i_{n(k)}$
($k=1,\ldots,2N-1$) with $n(k)=[k/w_c]$. The parameter $\alpha$
characterizes the ``strength'' of the perturbation, whereas
$w_c/2N=1/2N_c$ is the area of the perturbation cells.

A perturbed time step consists of first applying the unperturbed
time-evolution operator $B$, followed by a perturbation operator
$U_{\alpha;i_0,\ldots,i_{N_c-1}}$ with
$i_0,\ldots,i_{N_c-1}\in\{-1,1\}$ chosen at random, $\alpha$ being
fixed. We thus allow for a different perturbation at each step, in
contrast to Ref.~\cite{Peres1991} where a particular perturbed
evolution operator was applied repeatedly. After $n$ time steps, the
number of different perturbation
sequences---or histories---is $2^{nN_c}$.

Our specialization to vertically striped perturbation cells
involves no restriction relative to our work on the classical baker's
map. There we allowed for $(2^mw_c/2N) \times 2^{-m}$
rectangular perturbation cells, which are the
image of vertical stripes under $m$ applications of the baker's map,
$2^m \leq 2N/w_c = 2N_c$. Likewise, the perturbation operator for
``rectangular'' quantum-mechanical perturbation cells is
$U'=B^mUB^{-m}$ where $U$ is given by Eq.~(\ref{pert}). Using $U$ with
initial state $|\psi_0\rangle$ is equivalent to using $U'$ with
initial state $B^m|\psi_0\rangle$. The freedom to choose $m$---we use
$m=0$ for vertical stripes---is the same as the classical freedom to
choose the initial position of the ``decimal point'' in the symbolic
representation of the baker's map.

As a preliminary step, we show that perturbed evolution leads after
several steps to an ensemble of vectors that is similar to an ensemble
of vectors distributed randomly on Hilbert space.
A useful criterion for
determining the randomness of an ensemble of vectors $|\psi\rangle$ is
based on the moments of the quantity
\cite{Wootters1990}
\begin{equation}
W(|\psi\rangle)=-\sum_{i=0}^{2N-1} |\langle q_i|\psi\rangle|^2
\log(|\langle q_i|\psi\rangle|^2) \;,
\label{woot}
\end{equation}
which we call $W$-entropy to distinguish it from ordinary entropy
($\log$ denotes the binary logarithm, as throughout this paper). For
random vectors in $2N=16$-dimensional Hilbert space, the mean and
standard deviation of the $W$-entropy are given by $W=3.434\pm0.178$
bits, a result obtained on a computer by calculating $W(|\psi\rangle)$
for a large number of vectors $|\psi\rangle$ chosen at random from an
ensemble distributed uniformly over Hilbert space \cite{Wootters1990}.
(This mean value agrees with the exact formula for the mean value
\cite{Jones1990}, $W=[\Psi(2N+1)-\Psi(2)]/\ln2$, where $\Psi$ is the
digamma function.)  We compare the result for random vectors with the
first two moments of the $W$-entropy for an ensemble of vectors
created by the perturbed baker's map in a 16-dimensional Hilbert
space. Choosing  $2N_c=2$
and $\alpha=0.025=0.4/2N$, we create an ensemble of
approximately 20\,000 perturbed vectors by applying {\it different}
randomly chosen histories for $n=15$ perturbed time steps to the
initial vector $|\psi_0\rangle=|p_5\rangle$. We find for the first two
moments of the $W$-entropy the values $W=3.428\pm0.184$ bits, very
close to the moments for a random sample.

\begin{figure}
\epsffile{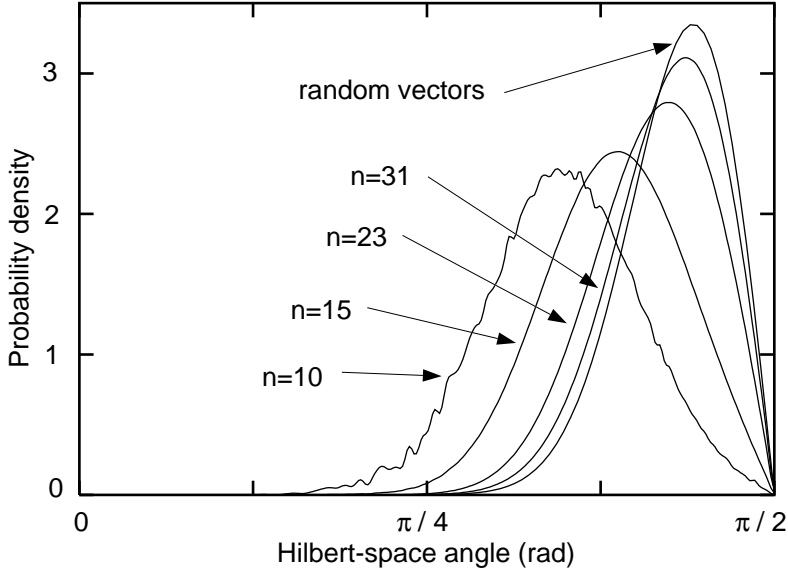}
\caption{Distribution of Hilbert-space angles for vectors
evolving under the perturbed quantum baker's map, shown for different
numbers of perturbed time steps $n$. For comparison, the distribution
for random vectors is also shown. The dimension of Hilbert space is
$2N=16$, the number of perturbation cells is $2N_c=2$, the
perturbation strength is $\alpha=0.025=0.4/2N$, and the initial vector
is $|\psi_0\rangle=|p_5\rangle$.}
\label{figdist}
\end{figure}

As a second check of randomness, we calculate the distribution of
Hilbert-space angles $\theta=\cos^{-1}(|\langle\psi'|\psi\rangle|)$
between vectors $|\psi\rangle$ and $|\psi'\rangle$ that have evolved
under the same perturbed quantum baker's map applied to the same
initial state as in the previous example. We compute the
Hilbert-space angle between each pair of vectors in each of three
ensembles of approximately 16\,000 vectors created by
applying different randomly chosen perturbation histories for 15, 23,
and 31 steps. In addition, we compute the Hilbert-space angle
between each pair of the $2^{10}$ vectors after 10 steps.  The
resulting distributions of Hilbert-space angles are displayed in
Fig.~\ref{figdist}.
After 31 steps the closest pair of vectors is
$27.1^\circ$ apart, a striking demonstration of the ``size'' of
16-dimensional Hilbert space, i.~e., of how many widely separated
vectors Hilbert space can accommodate even for a relatively small
dimension. For comparison, Fig.~\ref{figdist} also shows the
distribution $f(\theta)=30(\sin\theta)^{29}\cos\theta$ of
Hilbert-space
angles for a set of random vectors. These results show clearly
how the ensemble is randomized by the perturbed quantum baker's map.

\begin{figure}
\epsffile{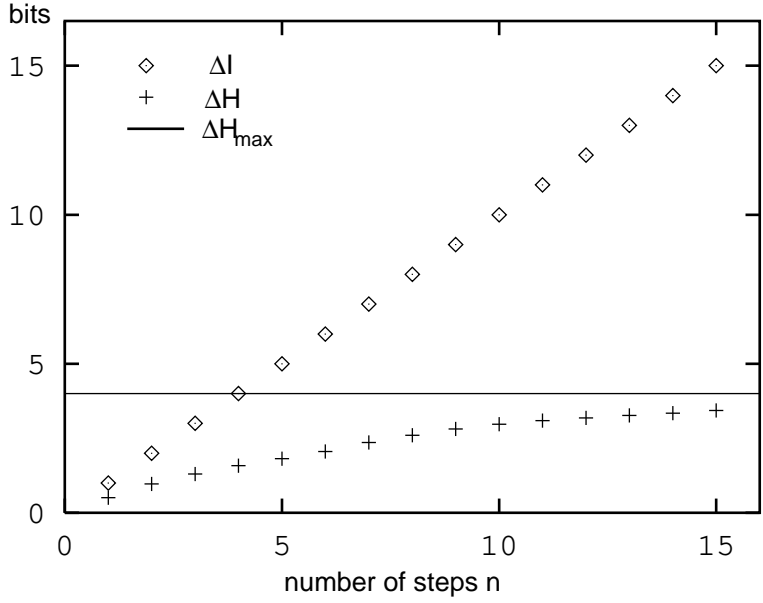}
\caption{Conditional algorithmic information $\Delta I$ needed to
track a vector evolving under the perturbed quantum baker's map,
compared to the increase in ordinary entropy $\Delta H$ that results
from averaging over the perturbation. The parameters are the same as
in Fig.~1.}
\label{dim16}
\end{figure}

We proceed now to compare the two strategies for extracting work
outlined above---coarse graining versus following the evolved vector
in fine-grained detail.  We estimate the conditional algorithmic
information $\Delta I$ needed---given background information---to
specify a typical perturbed vector after $n$ steps and compare it to
the increase in ordinary entropy $\Delta H$ that results from
averaging over the perturbation.  Our first example uses, as before, a
$2N=16$-dimensional Hilbert space, partitioned into $2N_c=2$
vertically striped perturbation cells. We choose a fixed perturbation
amplitude $\alpha=0.025=0.4/2N$ and an initial pure state
$|\psi_0\rangle=|p_5\rangle$, i.~e., a momentum eigenstate, which
corresponds to a horizontal stripe in the unit square. This
perturbation can be described completely by giving one bit per step,
to specify which of the two possible perturbation operators
$U_{\alpha;1}$ and $U_{\alpha;-1}$ is applied.  If the logarithmic
term \cite{Zurek1989b} that keeps track of the number of steps $n$ is
neglected, this sets an upper bound on the information $\Delta I$.
This upper bound is realized only if two different
histories of perturbed time steps always lead to two different
vectors at some level of resolution on Hilbert space. We choose a
resolution that regards two vectors as different if their
Hilbert-space angle exceeds $\delta\theta=\pi/50=3.6^\circ$
($|\langle\psi'|\psi\rangle|$ smaller than $0.998$). By comparing
numerically all possible histories, we find that, through 15 perturbed
time steps, {\it all\/} trajectories lead to distinguishable vectors.
Figure \ref{dim16} shows the resulting linear increase in the
information $\Delta I$.

\begin{figure}
\epsffile{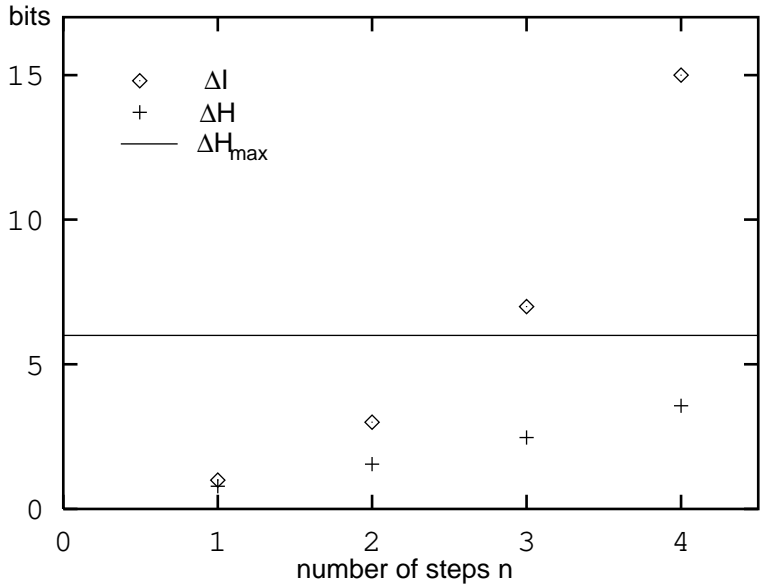}
\caption{$\Delta I$ and $\Delta H$ as in
Fig.~2, but with parameters $2N=64$, $2N_c=16$,
$\alpha=0.05\simeq3/2N$, and $|\psi_0\rangle=B^{-5}|p_1\rangle$.}
\label{dim64}
\end{figure}

Figure~\ref{dim16} also shows the ordinary entropy increase $\Delta
H$, obtained by determining the entropy of the density matrix that
results from averaging over all possible histories. It can be seen
that $\Delta I$ is always larger than $\Delta H$. Indeed, $\Delta H$
saturates at the value $\Delta H_{\rm max}=\log2N=4$ bits, the
logarithm of the dimension of Hilbert space, whereas $\Delta I$ is
only limited by $\Delta I_{\rm max}\simeq -2(2N-1)\log(\delta\theta/2)
\simeq 150$~bits, which is the logarithm of the number of different
vectors Hilbert space can accommodate \cite{Caves1993,Percival1992}.
Whereas $\Delta H_{\rm max}$ grows logarithmically with the dimension
$2N$ of Hilbert space, the maximum information $\Delta I_{\rm max}$
grows linearly with $2N$ and is enormous for macroscopic systems.

As in the classical case \cite{Schack1992}, the information $\Delta I$
grows more dramatically when the number of perturbation cells is
large. Figure~\ref{dim64} displays results for a 64-dimensional
Hilbert space with 16 vertically striped perturbation cells, each
containing four position eigenstates. The perturbation strength is
$\alpha=0.05\simeq3/2N$, and the initial state is $|\psi_0\rangle =
B^{-5}|p_1\rangle$, a state whose image under $B$ has negligible
support outside the leftmost perturbation cell. This means that, in
order to describe the perturbed state after the first time step, in
the perturbation operator $U_{\alpha;i_0,\ldots,i_7}$ only the sign
$i_0$ referring to the leftmost perturbation cell must be specified.
Since $B^2|\psi_0\rangle$, $B^3|\psi_0\rangle$, and
$B^4|\psi_0\rangle$ extend
over 2, 4, and 8 perturbation cells, we expect the number of
bits needed to specify the perturbed state to grow as
$\sum_{j=0}^{n-1} 2^j=2^n-1$ until the state extends over all
perturbation
cells. This behavior is verified in Fig.~\ref{dim64} using the
same method as for Fig.~\ref{dim16}.

Given these results and those of our previous paper \cite{Schack1992},
we have demonstrated similar hypersensitivity to perturbation in a
classically chaotic system and its quantum analogue. In both cases
the large information needed to track the perturbed evolution is due
to the large number of possible ways to perturb a state
\cite{Caves1993,Caves1993a}. In the
classical domain, chaos opens up the large space of
possibilities---phase-space patterns with structure on finer and finer
scales. Quantum mechanics operates inherently in an enormous space of
possibilities---the pure states on Hilbert space.
Hypersensitivity to perturbations means that more work can be
extracted by coarse graining than by following the perturbed evolution
in fine-grained detail. Our results provide a motivation for coarse
graining and thus an explanation of the second law of thermodynamics.

RS acknowledges the support of a fellowship from the Deutsche
Forschungsgemeinschaft.


\begin{thebibliography}{99}

\bibitem{QuantumChaos} See, e.~g., F. Haake,
{\em Quantum Signatures of Chaos\/} (Springer, New York, 1991);
{\em Quantum Chaos},
edited by H. A. Cerdeira, R. Ramaswamy, M. C. Gutzwiller, and G.
Casati
(World Scientific, Singapore, 1991);
{\em Quantum Chaos, Quantum Measurement},
edited by P. Cvitanovi\'c, I. Percival, and A. Wirzba
(Kluwer, Dordrecht, 1992).

\bibitem{Ford1991} J. Ford, G. Mantica, and G. H. Ristow,
Physica D {\bf 50}, 493 (1991).

\bibitem{Peres1992} A. Peres,
in {\em Quantum Chaos, Quantum Measurement},
edited by P. Cvitanovi\'c, I. Percival, and A. Wirzba
(Kluwer, Dordrecht, 1992), p.~249.

\bibitem{Schack1992} R. Schack and C. M. Caves,
Phys.\ Rev.\ Lett.\ {\bf 69}, 3413 (1992).

\bibitem{Arnold1968} V. I. Arnold and A. Avez, {\it Ergodic Problems
of Classical Mechanics\/} (Benjamin, New York, 1968).

\bibitem{Caves1993}
C.~M.\ Caves,
in {\it Physical Origins of Time Asymmetry},
edited by J.~J.\ Halliwell, J.~P\'erez-Mercader, and W.~H.\ Zurek
(Cambridge University Press, Cambridge, England, 1993).

\bibitem{Landauer1961}
R.~Landauer,
IBM J.\ Res.\ Develop.\
{\bf 5}, 183 (1961).

\bibitem{Landauer1988}
R.~Landauer,
Nature
{\bf 355}, 779 (1988).

\bibitem{Chaitin1987}
G.~J.~Chaitin,
{\it Information, Randomness, and Incompleteness\/}
(World Scientific, Singapore, 1987).

\bibitem{Zurek1989a}
W.~H.~Zurek,
Nature
{\bf 341}, 119 (1989).

\bibitem{Zurek1989b}
W.~H.~Zurek,
Phys.\ Rev.~A
{\bf 40}, 4731 (1989).

\bibitem{Balazs1989} N. L. Balazs and A. Voros,
Ann.\ Phys.\  {\bf 190}, 1 (1989).

\bibitem{Weigert1991} S. Weigert,
Phys.\ Rev.\ A {\bf 43}, 6597 (1991).

\bibitem{Ford1983} J. Ford,
Phys. Today {\bf 36}(4), 40 (1983).

\bibitem{Saraceno1990} M. Saraceno,
Ann.\ Phys.\  {\bf 199}, 37 (1990).

\bibitem{Peres1991} A. Peres,
in {\em Quantum Chaos},
edited by H. A. Cerdeira, R. Ramaswamy, M. C. Gutzwiller, and G.
Casati
(World Scientific, Singapore, 1991), p.~73.

\bibitem{Wootters1990} W. K. Wootters,
Foundations of Physics {\bf 20}, 1365 (1990).

\bibitem{Jones1990} K. R. W. Jones,
J. Phys.\ A {\bf 23}, L1247 (1990).

\bibitem{Percival1992} I. C. Percival,
in {\em Quantum Chaos, Quantum Measurement},
edited by P. Cvitanovi\'c, I. Percival, and A. Wirzba
(Kluwer, Dordrecht, 1992), p.~199.

\bibitem{Caves1993a} C. M. Caves, ``Information and entropy'',
submitted to Phys.\ Rev.\ E.

\end{thebibliography}
\end{document}